\documentclass[twocolumn,aps,floats,floatfix,nofootinbib,prd,superscriptaddress,tightenlines]{revtex4}
\usepackage{graphicx} \usepackage{bm} \usepackage{epsfig}
\usepackage{pstricks}

\begin{document}

\title{Post-Newtonian corrections to the motion of spinning bodies in NRGR}

\author{Rafael A. Porto}
\affiliation{Department of Physics, Carnegie Mellon University,
Pittsburgh, PA 15213}

\begin{abstract}
In this paper we include spin and multipole moment effects in the
formalism used to describe the motion of extended objects recently
introduced in \cite{nrgr}. A suitable description for spinning
bodies is developed and spin-orbit, spin-spin and quadrupole-spin
Hamiltonians are found at leading order. The existence of tidal,
as well as self induced finite size effects is shown, and the
contribution to the Hamiltonian is calculated in the latter. It is
shown that tidal deformations start formally at ${\cal O}(v^6)$
and ${\cal O}(v^{10})$ for maximally rotating general and compact
objects respectively, whereas self induced effects can show up at
leading order. Agreement is found for the cases where the results
are known.
\end{abstract}

\maketitle

\section{Introduction}

In a recent paper an Effective Field Theory (EFT) of gravity for
non spinning, spherically symmetric extended objects was
introduced \cite{nrgr}. Within the Post Newtonian (PN) \cite{pn}
framework this approach was coined NRGR (Non Relativistic General
Relativity) due to its similarities with EFT approaches to non
relativistic bound states in QED and QCD~\cite{ira}. However, the
EFT formalism can be applied to a variety of scenarios, for
instance the large small mass ratio case \cite{nrgr2}. NRGR is
relevant for understanding the gravitational power spectra emitted
by binary systems, an important class of candidate signals for
gravitational wave observatories such as LIGO or VIRGO
\cite{ligo,virgo}. The formalism allows for a clean separation of
the long wavelength gravitational dynamics from the details of the
internal structure. This separation enables us to calculate
corrections to all orders in the point particle approximation.
Furthermore, it was shown that the ambiguities \cite{v6,damour}
that plague the conventional PN calculations can be attributed to
the presence of higher-dimensional worldline terms in the action
whose coefficients
encode the short distance structure of the particles.\\

Building upon this idea, here we propose an extension of NRGR
which allows for the inclusion of spin and multipole moments. Spin
in General Relativity (GR) has been considered previously in the
literature from many different points of view (see for instance
\cite{tulc,spin,thorne1,dam} and references therein), and is
argued to play an important role in binary inspiral, particularly
for black holes \cite{will,will2}. Within the PN approximation
spin effects for binary systems have been calculated using
different techniques \cite{will,will2,kidd,cho,owen}. Dealing with
spinning objects in the point particle approximation inevitably
entails running into divergent integrals as one does in the non
spinning case. Regularization procedures, like Hadamard finite
part \cite{blanchet} or considering contributions from different
zones \cite{wise}, were invoked when dealing with point-like
sources \cite{owen}. However, it has been argued for instance for
the proposal of \cite{wise}, that the formalism is {\it
``considerably complicated but it is inevitable that we have to
adopt it to deal with divergences when we go to higher PN orders"}
\cite{cho}. As it has been repeatedly emphasized in \cite{nrgr}
that is not the case within an EFT approach. Here we will
explicitly see how a systematic, and consistent to all orders
approach is translated into our case as well.

The outline of the paper is as follows. In the first section we
review NRGR for non-spinning spherically symmetric objects
highlighting the main results. Then, we will generalize the
formalism to include internal angular as well as multipole moment
degrees of freedom extending the work of A. Hanson and T. Regge in
the realm of special relativity \cite{regge}. Afterwards we derive
the power counting and Feynman rules of NRGR and calculate the
leading spin-spin and spin-orbit potentials and show to reproduce
known results \cite{spin,will,will2}. A quadrupole-spin correction
to the gravitational energy is obtained for the first time (to my
knowledge) to leading order. The equivalence between different
choices for the spin supplementary condition is also shown.
Finally we discuss the insertion of non-minimal terms in the
worldline action and its relevance to renormalization. The
existence of two types of finite size effects encapsulated in a
new set of coefficients which can in principle be fixed by
matching to the full theory, is predicted: those which have a
renormalization group (RG) flow and naturally represent tidal spin
effects induced by the companion, and those that do not have a RG
flow and represent self induced effects, such as the spin induced
quadrupole moment due to the proper rotation of the objects
\cite{poisson,poisson2}. By power counting it is shown that
companion induced tidal effects start formally at 3PN, and 5PN for
maximally rotating general and compact objects respectively. Self
induced effects can show up at leading order. Details are
relegated to appendixes. We will study higher order PN
corrections, the radiative energy loss, matching and new possible
kinematic scenarios in future publications.

\section{NRGR}

In this section we will emphasize the main features of NRGR within
the PN formalism, detailed calculation and further references can
be found in the original proposal \cite{nrgr}.

\subsection{Basic philosophy}

The traditional approach to the problem of motion was introduced
by Fock \cite{Fock} who split it into two sub-problems. The {\it
internal problem}, which consists of understanding the motion of
each body around its center of mass, and the {\it external
problem} which determines the motion of the centers of mass of
each body. Decomposing the problem this way allows us to naturally
separate scales and henceforth calculate in a more systematic
fashion. The price one pays is the necessity of a {\it matching}
procedure which relies either in comparing with the full theory,
if known, or extracting unknown parameters from experiment. This
method is now called ``effective theory'', or EFT in the realm of
Quantum Field Theory (QFT), and has been used to great success in
many different branches of physics \cite{ira}. While at first
glance quantum field theoretical tools appear to introduce
unnecessary machinery for classical calculations\footnote{QFT
techniques have been recently used to calculate self-force effects
in a curved spacetime background \cite{Hu}.}, the power of the
method will be shown to reside in two facts: It allows for the
introduction of manifest power counting and naturally encapsulates
divergences into text book renormalization
procedures\footnote{Also bear in mind that the classical solution
is just the saddle point approximation to the path integral or
what is known as the ``tree level'' approximation.}. This means in
addition that in the EFT it is straightforward to calculate the
order at which a given term in
the perturbative series first contributes to a given physical observable.\\

Here we are going to tackle the problem of motion by treating
gravity coupled to point particle sources as the classical limit
of an EFT, i.e. the ``tree level approximation'', within the PN
formalism. Feynman diagrams will naturally show up as perturbative
techniques to iteratively solve for the full Green functions of
the theory. As it is known, GR coupled to distributional sources
is not generically well defined due to its non linear
character\footnote{I would like to thank Jorge Pullin for
discussions on this point.} \cite{wald}. This can be seen as a
formal obstacle to the PN expansion for point particle sources in
GR. Within an EFT paradigm this problem does not even arise since
one is not claiming to construct a full description to be
applicable to all regimes, but an effective theory which will
mimic GR coupled to extended objects within its realm of
applicability. In addition one can also argue that this EFT could
be seen as the low energy regime of a quantum theory of gravity
necessary to smear out point-like sources.\\

The idea of describing low energy quantum gravity as an EFT is not
new, for a review see \cite{eftg}. What makes NRGR appealing is
the uses of EFT to attack so called {\it classical} problems. QFT
has proven to be useful with classical calculations, as in
electromagnetic radiation where we can think of photons (QED) to
calculate a power spectrum. Here we will use the same idea
introducing ``gravitons'' as the quantum of the metric field which
will allow us to calculate the gravitational potential, from which
the equations of motion (EOM) are derived, as well as
gravitational radiation in a systematic fashion.

\subsection{Effective theory of extended objects}

The method of \cite{nrgr} is based in the explicit separation of
the relevant scales of the problem: the size of the objects $r_s$
(internal problem), the the size of the orbit $r$ (external
problem) and the natural radiation wavelength $r/v$, where $v\ll
c$ is the relative velocity in the PN frame. Finite size effects
are treated by the inclusion of a tower of new terms in the
worldline action which are needed to regularize the theory
\footnote{A similar approach can be found in \cite{Dphi} within
the realm of tensor-scalar theories. However, renormalization
issues as well as spin effects were left undiscussed.}. For a non
spinning spherically symmetric particle the most general action
consistent with the symmetries of GR is

\begin{equation}
\label{above} S=\int( - m + c_R R+c_V v^\mu v^\nu R_{\mu \nu}+...)
d\tau,
\end{equation}
where $R_{\mu\nu}$ and $R$ are the Ricci tensor and scalar
respectively. The series, involving higher order Riemann type
insertions, must be truncated within the desired accuracy to have
any predictive power. The coefficients of each of these new terms
can be determined by comparison with the full theory. In this case
the underlying theory is GR plus the internal equation of state of
the objects. The beauty of this method is that, since these are
1-body properties, we can match using any relevant observable, for
instance scattering processes, rather
than solving the complete problem of motion explicitly.\\

As it was shown in \cite{nrgr} the terms proportional to $c_R,c_V$
are generated by logarithmic divergences of the point particle
approximation. However it is possible to show that they are {\it
unphysical} in the sense that they can be removed from the
effective action by field redefinition (f.r.) and no trace is left
in observable quantities \cite{nrgr,damour}. Nevertheless from
here one concludes that not all divergences can be absorbed into
the mass and new counterterms are necessary. Furthermore, at
higher orders\footnote{$v^{10}$ for non spinning particles.} it
can be shown that (full Riemann dependent) finite size tidal
effects are induced which can not be removed from the theory. We
will see here that allowing the objects to spin also introduces
new terms in the worldline action. For the sake of completeness,
and given that the same idea will be used here later on, we will
sketch the reasoning. One starts by calculating the effective
action \cite{Peskin},

\begin{equation}
\Gamma[g_{\mu\nu}]={1\over m_{p}}\int  {d^4 k\over
(2\pi)^4}h_{\mu\nu}(-k) T_{(1p)}^{\mu\nu}(k)+\cdots,
\end{equation}
where $g_{\mu\nu}=\eta_{\mu\nu}+\frac{h_{\mu\nu}}{m_p}$, and
$h_{\mu\nu}$ the graviton field. Let us concentrate on the
contributions to the one point function $T_{(1p)}^{\mu\nu}(k)$. As
it was shown in \cite{nrgr} within dimensional regularization
techniques (dim. reg.), the logarithmic divergences in
$T_{(1p)}^{\mu\nu}(k)$ can not be absorbed into the mass and a new
counter-term of the form,

\begin{equation}
T_{ct}^{\mu\nu}(k)= (2\pi) \delta(k\cdot v)\left[c_R
(\eta^{\mu\nu} k^2 - k^\mu k^\nu) +{1\over 2} c_V k^2 v^\mu
v^\nu\right],
\end{equation}
is therefore needed. It is straightforward to conclude from here
the necessity of including two new terms in the effective action
as shown in (\ref{above}). Within dim. reg. an arbitrary mass
scale $\mu$ associated to the substraction point at which the
theory is renormalized is introduced. Given that the metric field
does not pick any anomalous dimension at tree level we must have
$\mu d\Gamma[g_{\mu\nu}]/d\mu=0$. Thus the explicit dependence on
the subtraction scale $\mu$ must be cancelled by allowing the
coefficients $c_{R,V}$ to vary with scale. The theory therefore
exhibits non-trivial {\it classical} RG scaling. As we are going
to show here spin dependent finite size effects are predicted by
similar arguments.

\subsection{The Post-Newtonian expansion}

Once the internal scale is taken into account by the introduction
of a series of new terms in the 1-body worldline action, the next
scale we have to integrate out is the orbit scale. In order to do
that we decompose the graviton field $h_{\mu\nu}$ into two pieces,

\begin{equation}
h_{\mu\nu}(x) = {\bar h}_{\mu\nu}(x) + H_{\mu\nu}(x)\label{hH},
\end{equation}
where $H_{\mu\nu}$ represents the off-shell potential gravitons,
with
\begin{eqnarray}
\partial_i H_{\mu\nu}\sim {1\over r} H_{\mu\nu} & &  \partial_0 H_{\mu\nu}\sim {v\over r} H_{\mu\nu} ,
\end{eqnarray}
and ${\bar h}_{\mu\nu}$ describes an on-shell radiation field
\begin{equation}
\partial_\alpha {\bar h}_{\mu\nu} \sim {v\over r} {\bar h}_{\mu\nu}.
\end{equation}

We can now further decompose $H_{\mu\nu}$ by removing from it the
large momentum fluctuations,

\begin{equation}
H_{\mu\nu}(x) = \int \frac{d^3{\bf k}}{(2\pi)^3} e^{i{\bf k}\cdot
{\bf x}} H_{{\bf k}\mu\nu}(x^0).
\end{equation}

The advantage of this redefinition is that now derivatives acting
on any field in the EFT scale in the same way, $\partial_\mu\sim
v/r$, so it is easy to count powers of $v$ coming from derivative
interactions.\\

\begin{figure}[!t]
\centerline{\scalebox{0.6}{\includegraphics{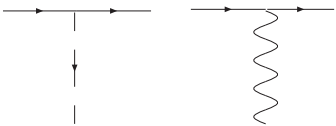}}} \vskip-0.3cm
\caption[1]{Leading order mass vertex. the dashed line represents
a potential graviton, whereas the wavy line stands for the full
graviton propagator.}\label{mH}
\end{figure}

The effective {\it radiation} NRGR Lagrangian, with the potential
gravitons integrated out, can then be derived by computing the
functional integral,

\begin{equation}
\label{eq:NRGRPI} \exp[i S_{NRGR}[x_a,{\bar h}]] = \int {\cal
D}H_{\mu\nu} \exp[i S[{\bar h}+H,x_a] + i S_{GF}],
\end{equation}
where $S_{GF}$ is a suitable gauge fixing term.
Eq.~(\ref{eq:NRGRPI}) indicates that as far as the potential modes
$H_{\mu\nu}$ are concerned ${\bar h}_{\mu\nu}$ is just a slowly
varying background field.  To preserve gauge invariance of the
effective action, we choose $S_{GF}$ to be invariant under general
coordinate transformations of the background metric ${\bar
g}_{\mu\nu}(x) = \eta_{\mu\nu} + {\bar h}_{\mu\nu}(x)$. This whole
procedure is what is usually known as the ``Background Field
Method'', originally introduced by DeWitt \cite{DeWitt} in
canonical quantum gravity and used by t'Hooft and Veltman for the
renormalization of gauge theories \cite{thooft}. By expanding the
Einstein-Hilbert action using (\ref{hH}) we can immediately read
off Feynman rules \cite{nrgr}. For potential gravitons, which we
are going to represent by a dashed line, the propagator is given
by,

\begin{equation}
\langle H_{{\bf k}\mu\nu} (x^0)H_{{\bf q}\alpha\beta}(0)\rangle =
-(2\pi)^3\delta^3({\bf k} + {\bf q}){i\over {\bf
k}^2}\delta(x_0)P_{\mu\nu;\alpha\beta},
\end{equation}
where $P_{\mu\nu;\alpha\beta} = {1\over 2}\left[\eta_{\mu\alpha}
\eta_{\nu\beta} + \eta_{\mu\beta} \eta_{\nu\alpha} - \eta_{\mu\nu}
\eta_{\alpha\beta}\right]$. The radiation gravitons, which will be
represented by a curly line, have the usual spin 2 massless
propagator. A wavy line will be used for the full propagator. We
also need to consider mass insertions which will just provide a
vertex interaction \cite{nrgr},

\begin{equation}
\sum_a\frac{m_a}{m_p} \left[{1\over 2} h_{00} + h_{0i} {{\bf
v}_a}_i  + {1\over 4} h_{00} {\bf v}^2_a + {1\over 2} h_{ij} {{\bf
v}_a}_i {{\bf v}_a}_i \right]+ \cdots,
\end{equation}
where $h_{00},h_{0i},h_{ij}$ are evaluated on the point particle
worldline (the leading order graviton-mass vertex is shown in
fig.~\ref{mH}). Following standard power counting procedures we
arrive to the scaling laws for the NRGR fields shown in table I
\cite{nrgr,ira}. In the last column we have introduced
$m^2_p=\frac{1}{32\pi G_N}$ the Planck mass and $L=mvr$ the
angular momentum.

\begin{table}
\begin{eqnarray}
\nonumber
\begin{array}{c|c|c|c}
  {\bf k} & H^{\bf k}_{\mu\nu} & {\bar h}_{\mu\nu}  & m/m_p\\
\hline
   1/r      & r^2 v^{1/2}                   & v/r & \sqrt{L v}\\
\end{array}
\end{eqnarray}
\caption{NRGR power counting rules.} \label{tab}
\end{table}

The effective action in (\ref{eq:NRGRPI}) will be a function of
the worldline particles (treated as external sources) and the
radiation field which allows us to calculate the energy loss due
to gravitational radiation as well as the gravitational binding
potential from which the EOM are obtained. To get $S_{eff}(x^a)$
we simply integrate out the radiation field ${\bar h}$.
$S_{eff}(x^a)$ has a real part which represents the effective
potential for the 2-body system\footnote{Remember we are treating
the worldline of the particles as external sources, namely
$x^a\equiv J$, therefore $S_{eff}$ is just the partition function
$e^{iS_{eff}}\equiv <0|0>^J \sim e^{i\Gamma(J,\dot J...)T}$, with
$\Gamma(J)$ the effective action for the sources as $T \to \infty$
\cite{coleman}.}. It also has an imaginary part that measures the
total number of gravitons emitted by a given configuration
$\{x_\mu^a\}$ over an arbitrarily large time $T\rightarrow\infty$,

\begin{equation}
{1\over T} \,\mbox{Im} S_{eff}(x_a) =  {1\over 2} \int dE d\Omega
{d^2\Gamma\over dE d\Omega},
\end{equation}
where $d\Gamma$ is the differential rate for graviton emission
from the binary system from which the power spectrum is computed.

\section{Internal degrees of freedom I: Spinning particle}

Here we will follow closely the ideas developed in \cite{regge}.
We will start formulating a Lagrangian formalism to deal with
internal angular as well as multipole moment degrees of freedom
which will enable us to describe a richer tensor structure. We
will introduce the basic elements first, then we will construct
the action and show how to reproduce Papapetrou equations for
spinning particles in GR \cite{papa}. The issue of constraints,
the correct number of degrees of freedom and the
angular-velocity/spin relationship will be discussed at the end of
the section.

\subsection{Basics}

Given a spacetime structure $(g,M)$ we can always find at each
point $x \in M$ a coordinate system where the metric looks locally
flat at the point. Such a transformation can be expressed as:

\begin{eqnarray}
\label{orth} \eta_{IJ}=e^{\mu}_Ie^{\nu}_J g_{\mu\nu},\\
\eta^{IJ}e^{\mu}_Ie^{\nu}_J = g^{\mu\nu},\label{tetr}
\end{eqnarray}
with $\eta_{IJ} \equiv (1,-1,-1,-1)$ the Minkowski metric and
$e^{\mu}_I$ a set of $I=0..3$ orthonormal basis vectors such that
the tensor metric is diagonalized at the point. From now on
capital Latin letters will denote internal indexes (notice that
they transform in $SO(3,1)$ due the residual Lorentz invariance),
the other conventions are as usual. Given a tetrad we can define
its transport through the particle's worldline using Fermi-Walker
ideas as \cite{mtw},

\begin{equation}
{\dot e}^I_{\mu}\equiv \frac{D e_I^{\mu}}{d\lambda}=
u^{\alpha}\nabla_{\alpha}e^{\mu}_I=-\Omega^{\mu\nu}e_{I\nu},
\label{t}
\end{equation}
where $\nabla_{\alpha}$ is the covariant derivative compatible
with $g$, namely $\nabla_{\alpha}g_{\mu\nu}\equiv
g_{\mu\nu;\alpha}=0$, and $\Omega^{\mu\nu}$ is an antisymmetric
tensor which therefore preserves (\ref{orth}). One can invert the
previous relation using (\ref{orth}),

\begin{eqnarray}
\Omega_{\beta\alpha}&=&\eta^{IJ}e_{\beta I} \frac{De_{\alpha
J}}{d\tau}\nonumber\\&=& \left(\frac{de_{\alpha
J}}{d\tau}-\Gamma^\sigma_{\alpha\gamma}e_{\sigma J}
u^{\gamma}\right)\eta^{IJ}e_{\beta I}\label{Ome}.
\end{eqnarray}

Notice that (\ref{Ome}) implies the antisymmetry directly from $g_{\nu\mu;\mu}=0$.\\

The introduction of $e^{\mu}_I$ is equivalent to adding an element
of $SO(3,1)$ to the worldline of the particle to describe
rotations \cite{regge}. Following these ideas we will therefore
construct an action in terms of the generalized coordinates and
velocities $(x^{\mu},u^{\nu},e^I_{\mu}, {\dot e}^I_{\mu})$. The
number of degrees of freedom in $e^\mu_I$ is 3 more than we need
to describe 3-rotations. We will see however that we can impose a
set of kinematic constraints which will ensure the correct number.

\subsection{Action principle and EOM}

So far we have characterized the extra degrees of freedom we need
in order to construct a Lagrangian for the spinning particle. In
the process to construct the action we will demand in addition to
general covariance, internal Lorentz invariance as well as
reparametrization invariance (RPI). This will naturally restrict
ourselves to Lagrangians of the form
$L(x^{\mu},u^{\nu},\Omega^{\mu\nu})$. It is however natural,
instead of using $\Omega^{\mu\nu}$ as coordinates to treat them as
velocities of angular degrees of freedom which will lead us to a
natural interpretation of spin. It is easy to see there are four
different scalar quantities (neglecting parity violating terms) we
can consider (schematically),

\begin{eqnarray}
a_1&=&u^2\\
a_2&=&\Omega^2\\
a_3&=&u\Omega\Omega u\\
a_4&=&\Omega\Omega\Omega\Omega
\end{eqnarray}
where contractions are made with the space-time metric
$g_{\mu\nu}$. Using these quantities our Lagrangian will be in
principle a general expression of the form $L(a_1,a_2,a_3,a_4)$.
We will neglect multipole moments throughout this section the
inclusion of which will be studied later on. The objects will be
therefore considered symmetric with respect to their rotational axis.\\

In order to introduce the idea of spin we will define the
antisymmetric tensor $S^{\mu\nu}$ and momentum $p^{\mu}$ by,

\begin{equation}
\delta L = -p^{\mu}\delta u_{\mu} -
\frac{1}{2}S^{\mu\nu}\delta\Omega_{\mu\nu},
\end{equation}
where the minus sign corresponds to the correct non relativistic
limit \cite{regge}. From these definitions we will have,

\begin{eqnarray}
p^{\alpha}&=& -2u^{\alpha}\frac{\partial L}{\partial
a_1}-2\Omega^{\alpha\nu}\Omega_{\nu\rho}u^{\rho}
\frac{\partial L}{\partial a_3}\label{11}\\
S^{\mu\nu}&=& -4\Omega^{\mu\nu}\frac{\partial L}{\partial
a_2}-2(u^{\mu}\Omega^{\nu\lambda}u_{\lambda}-
u^{\nu}\Omega^{\mu\lambda}u_{\lambda})\frac{\partial L}{\partial
a_3}\nonumber
\\&-& 8\Omega^{\nu\beta}\Omega_{\beta\alpha}\Omega^{\alpha\mu}\frac{\partial
L}{\partial a_4}.\label{12}
\end{eqnarray}

The variation of the action consists of two pieces. Let us
concentrate first in the tetrad part. Using the definition of spin
we will have to deal with,

\begin{widetext}
\begin{equation}
\delta S = -\int d\tau
S^{\alpha\beta}\delta{\Omega_{\alpha\beta}}=-\int d\tau
\left(-\frac{DS^{\alpha\beta}}{D\tau}e_{K\alpha}-\frac{De_{K\alpha}}{D\tau}S^{\alpha\beta}+
S^{\rho\nu}\frac{De_{J\nu}}{D\tau}e_{\rho K}
e^{J\beta}\right)\delta e^K_{\beta}.
\end{equation}
\end{widetext}

The equation of motion can be directly read from the above
expression, multiplying by $e^{K\mu}$ we get (using (\ref{Ome})),

\begin{equation}
\frac{DS^{\mu\nu}}{D\tau}=
S^{\mu\lambda}{\Omega_{\lambda}}^{\nu}-{\Omega^{\mu}}_{\lambda}S^{\lambda\nu}=p^{\mu}u^{\nu}-u^{\mu}p^{\nu}\label{eq1},
\end{equation}
where the last equality follows from (\ref{11},\ref{12})
\cite{regge}. Notice that we have not specified a Lagrangian up to
this stage. It is easy to see from (\ref{eq1}) that,

\begin{equation}
\frac{DS^{IJ}}{D\tau}\equiv
\frac{D(S^{\alpha\beta}e^I_{\alpha}e^J_{\beta})}{D\tau}=0\label{eq1n},
\end{equation}
which shows that spin projected with respect to the $e^I_{\alpha}$
frame remains constant. In addition one can also show that the
scalar $S^2\equiv \frac{1}{2}S^{\mu\nu}S_{\mu\nu}$ is conserved.
As a further property it is also instructive to notice that
$S^{\mu\nu}S^{*}_{\mu\nu}$ is also a constant of the motion, where
$S^{*}_{\mu\nu}=\frac{1}{2}\epsilon_{\mu\nu\alpha\beta}S^{\alpha\beta}$.\\

In order the get the $\delta x$ piece of $\delta S$, a shortcut
can be taken by going to a locally flat coordinate system where
the connection terms are zero at the point. Promoting the
derivatives to covariant ones in this frame we will end up with,

\begin{eqnarray}
\frac{Dp_{\gamma}}{D\tau}&=&\frac{1}{2}S^{\alpha\beta}\left(\Gamma_{\alpha\beta\sigma
,\gamma}-\Gamma_{\alpha\beta\gamma ,\sigma}
\right)u^{\sigma}=\nonumber\\&=&-{1\over
2}R_{\gamma\sigma\alpha\beta}S^{\alpha\beta}u^{\sigma},\label{eq2}
\end{eqnarray}
where we have used the form of the Riemann tensor in a locally
flat coordinate system. Written this way we can promote now the
equation to all reference systems since it is covariant. We
therefore recognize in (\ref{eq1},\ref{eq2}) the well known
Papapetrou equations \cite{papa}. Remarkably, although in terms of
the tetrad these equations depend on the choice of Lagrangian, as
a function of spin and momentum the evolution equations are
action-independent as far as curvature terms are not inserted.
Including curvature terms in the effective action will be relevant
to introduce finite size effects and will modify these equations
for extended objects.

\subsection{Constraints and angular-velocity/spin relationship}

In order to describe the correct number of degrees of freedom we
need to add a set of constraints to the EOM (\ref{eq1},\ref{eq2}).
A well defined angular-velocity/spin relationship is also
necessary to extend our power counting rules to the spinning case.
We will show here that both features are related. Here we will
closely follow \cite{regge}, to which we refer the reader for
details, other approaches may be found in \cite{pryce}. This
section relies on a basic knowledge of constrained systems, for
further
details see \cite{dirac,teit}.\\

It is natural to impose the following (covariant) constraints in
the space of solutions,

\begin{equation}
V^{\mu}=S^{\mu\nu}p_{\nu} \approx 0,\label{const}
\end{equation}
where just three of the four components are independent, and
``$\approx$" stands for weakly vanishing \cite{dirac}. This set of
constraints are second class, namely they have non vanishing
Poisson bracket among themselves, and therefore reduce the number
of degrees of freedom from $6$ to $3$ $SO(3)$ parameters as
expected. It can be shown in addition there
is a Lagrangian from which (\ref{const}) kinematically follows \cite{regge}.\\

We need to guarantee the constraints in (\ref{const}) are
preserved upon evolution. It is however possible to show from
(\ref{eq1},\ref{eq2}) that $\frac{DV^{\mu}}{d\tau} \approx 0$ will
be satisfied provided

\begin{equation}
p^{\alpha}= m
u^{\alpha}-\frac{1}{2m}R_{\beta\nu\rho\sigma}S^{\alpha\beta}S^{\rho\sigma}u^{\nu},\label{up}
\end{equation}
with $m^2(S^2)\equiv p^2$ defined by (\ref{11},\ref{12}). This
also means that the difference between $p^{\nu}/m$ and $u^{\nu}$
is higher order in the PN expansion and we can consider
$S^{\mu\nu}u_{\nu}=0$ as well as $\frac{dS^{\mu\nu}}{d\tau}=0$ to
leading order.\\

It is possible to show that (\ref{const}) implies
$C_1=S^{\mu\nu}S^{*}_{\mu\nu} \approx 0$, which is a first class
constraint. There is therefore, in addition to RPI
($C_2=p^2-m(S^2)\approx 0$), a gauge freedom which can be
attributed to the choice of the temporal vector of the tetrad
$e^0_{\mu}$, and a sensible choice of gauge is then
$\psi^\mu=e^\mu_0-p^\mu/m \approx 0$ \cite{regge}. This gauge,
jointly with (\ref{const}), also translates into a choice of
center of mass of the object \cite{spin,will2}, and implies as
well $\Omega^{\mu\nu}p_{\nu}=\frac{Dp^{\mu}}{D\tau}$, from which
we get

\begin{equation}
\Omega_{\mu\nu} \sim S_{\mu\nu} -
\frac{1}{2m}R_{\mu\nu\alpha\beta}S^{\alpha\beta} + RRSSS + ...
\label{omegas},
\end{equation}
where we have used (\ref{const},\ref{up}). We can indeed obtain
the angular-velocity/spin relationship by matching the evolution
equation for the tetrad in a Minkowski background obtaining (see
appendix A),

\begin{equation}
S^{\mu\nu}= \frac{I}{(u^2)^{1/2}}(\Omega^{\mu\nu}
+\frac{I}{2m}R_{\mu\nu\alpha\beta}\Omega^{\alpha\beta}+...)\label{omegas2},
\end{equation}
with $I$ the moment of inertia, and a particular Lagrangian is
chosen to ensure (\ref{const}) \cite{regge}. The main results of
this section are therefore equations (\ref{up},\ref{omegas2}),
from which we conclude that in a theory where (\ref{const}) is
kinematically imposed spin and angular-velocity are naturally
related and proportional in flat space. Within an EFT approach,
these relationships are all we need to construct the NRGR
extension for spinning bodies.

\section{NRGR for spinning bodies}

\subsection{Power counting}

The power counting rules in NRGR have been developed in
\cite{nrgr}. Here we are going to extend them to include spin
degrees of freedom. As we shall show the only necessary change is
the inclusion of spin insertions at the vertices.

First of all notice that, from the constraints at leading order,

\begin{equation}
S^{\mu\nu}u_{\nu}=0\to S^{j0}=S^{jk}u^k,
\end{equation}
which implies a suppression of the temporal components with
respect to the spatial ones. In other words spin is represented by
a 3-vector, $S^k=\frac{1}{2}\epsilon^{kij}S_{ij}$, in the rest
frame of the particle as expected. We will concentrate here in
compact objects like neutron stars or black holes where the
natural length scale can be taken to be their Schwarzschild radius
$r_s \sim Gm$ (for general objects see appendix B), and hence a
momentum of inertia scaling as $I\sim m^3/m^4_p$. For the spin
angular momentum we will have,
\begin{equation}
S=I\omega=I\frac{v_{rot}}{r_s}\sim mv_{rot}r_s < mr_s \sim Lv.
\end{equation}

We therefore see that spin gets suppressed with respect to the
orbital angular momentum, even for the maximally rotating case
($v_{rot}=1$). We can also assume a different kinematical
configuration with the particles co-rotating, namely
$\frac{v_{rot}}{r_s}=\frac{v}{r}$. In the former $S \sim
\frac{mvr^2_s}{r}=L\frac{r^2_s}{r^2}\sim Lv^4$. We will generally
power count spin as
\begin{equation}
S \sim Lv^s,
\end{equation}
with $s=1$ and $s=4$ in the maximally rotating and co-rotating scenario respectively.\\

To obtain the scaling laws we have assumed the usual
proportionality at leading order between spin and angular velocity
which was obtained from (\ref{omegas2}). As usual subleading
scalings will be naturally taken into account by the insertion of
higher order terms in the worldline action. What we have learnt
here is that spin effects are in any case subleading in the PN
expansion, and the scaling laws developed in \cite{nrgr} still
hold and spin contributions can be treated as a perturbation.

\subsection{Feynman rules: spin-graviton vertex}

To construct the effective theory for gravitons we need to expand
the metric around a Minkowski background, namely
$g_{\alpha\beta}=\eta_{\alpha\beta}+\frac{h_{\alpha\beta}}{m_p}$.
There is however a subtle point in doing this given that
(\ref{tetr}) leads to,

\begin{eqnarray}
\eta_{IJ}e_{\mu}^Ie_{\nu}^J&=&
\eta_{\mu\nu}+\frac{h_{\mu\nu}}{m_p},\label{he1}\\
e_{\mu}^I = \Lambda^I_{\mu}+\delta e_{\mu}^I\to \delta
e^I_{\mu}&=&\frac{1}{2m_p} h_{\mu\nu}\Lambda^{\nu I}+...\nonumber
\end{eqnarray}
where $\Lambda^{I\alpha}$ is an element of the Lorentz group. We
will also need $\delta e^{\mu I}$ which is defined through the
inverse metric \cite{velt},

\begin{eqnarray}
g^{\mu\nu}&=&\eta^{\mu\nu}+h^{\mu\nu},\nonumber\\
h^{\mu\nu}&=&\eta^{\mu\mu'}\eta^{\nu\nu'}
(-h_{\mu'\nu'}+h_{\mu'}^{\alpha}h_{\alpha\nu'}-...), \label{eup}
\end{eqnarray}
where indexes are raised with $\eta^{\mu\nu}$. We will therefore
have,

\begin{equation}
\delta
e^{I\mu}=-\frac{1}{2m_p}\eta^{\mu\nu}h_{\nu\alpha}\Lambda^{\alpha
I}+...,\label{he2}
\end{equation}

\begin{figure}[!t]
\centerline{\scalebox{0.6}{\includegraphics{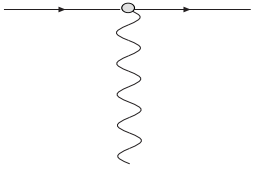}}}
\vskip-0.3cm \caption[1]{Leading order spin-graviton vertex
interaction. The blow represents an spin insertion.}\label{gsv}
\end{figure}

One can immediately see how to proceed by comparison with what it
has been done in \cite{regge} within flat space where the angular
velocity was defined as,

\begin{equation}
\Omega^{\mu\nu}_M \equiv
\Lambda^{\mu\alpha}\frac{d\Lambda^{\nu}_{\alpha}}{d\tau},
\label{omM}
\end{equation}
with $\Lambda^{\mu\nu}$ describing the {\it rotation} of the
particle, and $M$ stands for Minkowski. Expanding the action using
(\ref{he1},\ref{he2}) we will thus obtain the spin-graviton
interaction to all orders in a flat space background as a function
of the graviton field and (\ref{omM}) (see appendix C for
details). To leading order in the weak field expansion(see
fig.~\ref{gsv}),

\begin{equation}
L_0 =
\frac{1}{2m_p}h_{\alpha\gamma,\beta}S^{\alpha\beta}_Mu^{\gamma},\label{sg}
\end{equation}
where $S^{\mu\nu}_M=-\frac{\partial L}{\partial \Omega_{\mu\nu}}$
at $g_{\mu\nu}=\eta_{\mu\nu}$ (we will drop the $M$ from now on).
The expression in (\ref{sg}) is remarkably action independent if
written in terms of spin and the graviton field. The Lagrangian
dependence enters in the so far unknown function $S(\Omega)$. This
conclusion however just applies to the leading term and a choice
of action is necessary to obtain the Feynman rules to all orders.
At next to leading order in the weak gravity limit we will have,

\begin{equation}
L_1 = \frac{1}{4m^2_p}S^{\beta\gamma}u^{\mu}h_\gamma^\lambda
({1\over 2}h_{\beta\lambda,\mu}+
h_{\mu\lambda,\beta}-h_{\mu\beta,\lambda}),\label{h2sg}
\end{equation}
where a particular action has been chosen to ensure (\ref{const}).
A different choice of Lagrangian will imply different Feynman
rules. However, bear in mind that different actions will differ in
the spin/angular-velocity relationships and might not lead
kinematically to (\ref{const}). The physics will be invariant once
these differences are taken into account.

To calculate in the EFT we need to match (\ref{sg},\ref{h2sg})
into NRGR using the power counting rules developed in \cite{nrgr}
plus the spin insertions. Up to 2PN, for maximally rotating
compact objects we will get for the spinning part of the NRGR
Lagrangian,

\begin{eqnarray}
L^{NRGR}_{1PN} &=& \frac{1}{2m_p}H_{i0,k}S^{ik},\label{sgnr1}\\
L^{NRGR}_{1.5PN} &=& \frac{1}{2m_p}\left(H_{ij,k}S^{ik}u^j + H_{00,k}S^{0k}\right),\label{sgnr15}\\
L^{NRGR}_{2PN} &=& \frac{1}{2m_p}\left(H_{0j,k}S^{0k}u^j +
H_{i0,0}S^{i0}\right)\nonumber
\\ &+& \frac{1}{4m^2_p}S^{ij}\left(H^{\lambda}_j H_{0\lambda,i} -
H^k_j H_{0i,k}\right).\label{sgnr2}
\end{eqnarray}

The procedure follows systematically as shown in appendix C.

It is an useful exercise to check the gauge invariance of
(\ref{sg}), or in other words to obtain the leading stress energy
tensor. It is straightforward to calculate
$T_{(1)}^{\mu\nu}=-\frac{\partial L}{\partial g_{\mu\nu}}$ at
$g_{\mu\nu}=\eta_{\mu\nu}$ getting,

\begin{equation}
T_{(1)}^{\mu\nu}=-\frac{1}{2}\partial_{\beta}
\left(S^{\beta\mu}u^{\nu}+S^{\beta\nu}u^{\mu}\right),\label{st}
\end{equation}
which agrees at zero order with the original proposal of Dixon
\cite{dixon} and also Bailey and Israel (BI) \cite{israel} (see
appendix D).

The Ward identity,

\begin{equation}
\partial_{\mu}T_{(1)}^{\mu\nu} \sim
\partial_{\beta}\frac{dS^{\beta\nu}}{d\tau}=0\label{gt},
\end{equation}
is therefore obeyed since spin is constant to leading order in the
PN expansion.

\subsection{Leading order graviton exchange}

\begin{figure}[!t]
\scalebox{0.6}{\includegraphics{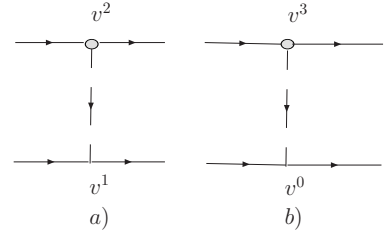}}\vskip-0.3cm
\caption[1]{Leading order spin-orbit interaction. Diagram a) takes
into account a $v^1$ mass insertion. Diagram b) correspond to the
$v^3$ spin-graviton vertex.}\label{sso}
\end{figure}

\begin{figure}[!t]
\scalebox{0.6}{\includegraphics{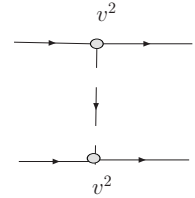}} \vskip-0.3cm
\caption[1]{Leading order spin-spin interaction.}\label{ss}
\end{figure}

Our goal from now on is to calculate the leading order piece (one
graviton exchange) of the potential energy due to spin-orbit
(fig.~\ref{sso}) and spin-spin (fig.~\ref{ss}) couplings coming
from (\ref{sgnr1},\ref{sgnr15})\footnote{Self energy terms are not
considered since they yield scaleless integrals.}. The leading
order spin-orbit contribution is the sum of two pieces,

\begin{widetext}
\begin{equation}
fig.~\ref{sso} = \frac{-i m_2}{4m_p^2}\int dt dt'
\frac{d^3p}{(2\pi)^3}
\partial_\beta
\left(\frac{e^{-i\vec{p}(\vec{x}(t)-\vec{y}(t'))}}{\vec{p}^2}\delta(t-t')
P_{0\epsilon;\alpha\gamma}\right)S_1^{\beta\alpha}u_1^\gamma(t)u_2^{\epsilon}(2-\delta^\epsilon_0).
\end{equation}
\end{widetext}

We need to distinguish two different cases: the temporal and
spatial derivative. The temporal derivative will just hit
$\delta(t-t')$ and can be integrated by part bringing down a
velocity factor. To leading order we would need to consider
$\epsilon=\gamma=0$. However, $P_{00;\alpha 0}=0$ unless
$\alpha=0$, which leads to a term proportional to $S^{00}=0$.
There is then no contribution from the temporal derivative and we
just need to concentrate in the spatial part and the terms,

\begin{equation}
-i m_2G_N\partial_j\int dt
\frac{1}{|\vec{x}(t)-\vec{y}(t)|}P_{0\epsilon;\alpha\gamma}S_1^{j\alpha}u_1^\gamma(t)u^\epsilon_2(t)(2-\delta^\epsilon_0).
\end{equation}

Notice that we have three possible contributions, one coming from
$\epsilon=0,\alpha=l,\gamma=k$, another where
$\epsilon=l,\alpha=k,\gamma=0$ and finally
$\epsilon=\alpha=\gamma=0$. The latter looks at first as a $v^0$
piece, however this is misleading since our spin choice implies
$S^{j0}=S^{jl}u^l$. Adding all the terms one gets,

\begin{equation}
fig.~\ref{sso} = i\int dt
\frac{-2G_Nm_2}{|\vec{x}(t)-\vec{y}(t)|^2}\left((\vec{n}\times
\vec{u_1})\cdot \vec{S_1}-(\vec{n}\times \vec{u_2})\cdot
\vec{S_1}\right),
\end{equation}
where $\vec{n}$ is the unit vector in the $(\vec{x}-\vec{y})$
direction and we have used $S^{kj}=\epsilon^{kji}S^i$. Joining the
mirror image we will end up with,

\begin{equation}
V_{SO}=\frac{2G_N}{r^2}\mu(\vec{n}\times \vec{v})\cdot
\left(\left(1+\frac{m_1}{m_2}\right)\vec{S_2}+\left(1+\frac{m_2}{m_1}\right)\vec{S_1}\right)\label{Eso},
\end{equation}
for the spin-orbit potential, where $\mu$ is the reduced mass,
$r=|\vec{x}(t)-\vec{y}(t)|$ and $\vec{v} \equiv \vec
{u}_1-\vec{u}_2$.

Let us now consider the spin-spin interaction. The leading order
contribution is (see fig.~\ref{ss}),

\begin{equation}
\frac{i}{4m_p^2} \partial_{y_k}\partial_{x_{k'}}\int dt
\frac{d^3p}{(2\pi)^3}
e^{i\vec{p}(\vec{x}(t)-\vec{y}(t))}\frac{1}{\vec{p}^2}
P_{j'0;j0}S_1^{jk}S_2^{j'k'}.
\end{equation}

Using $P_{j'0;j0}=-\frac{1}{2}\delta_{jj'}$ it is straightforward
to show
\begin{equation}
V_{SS}= -\frac{G_N}{r^3}\left(\vec{S}_1\cdot
\vec{S}_2-3\vec{S}_1\cdot
\vec{n}\vec{S}_2\cdot\vec{n}\right)\label{Ess},
\end{equation}
for the spin-spin binding potential. It is easy to see by power
counting that $V_{SO}dt \sim Lv^3$ and $V_{SS}dt \sim Lv^4$,
effectively 1.5PN and 2PN for maximally rotating compact objects.

By comparison with the results in \cite{spin}, it is immediate to
notice there is a mismatch in the spin-orbit contribution
(\ref{Eso}), which can be traced back to the choice of spin
supplementary condition in (\ref{const}). As it was noticed in
\cite{spin,dam,kidd,will} this discrepancy is associated to the
choice of center of mass of each body. It can be shown there is a
coordinate transformation that relates the center of mass choice
which follows from (\ref{const}) and the so called baryonic
coordinates (implicitly used in \cite{spin}), where the center of
mass is defined through the baryonic density, and one has
$S^{i0}=\frac{1}{2} S^{ij}u^j$. Had we calculated the spin-orbit
term within the baryonic condition we would obtain,

\begin{equation}
{\bar V}_{SO}=\frac{2G_N}{r^2}\mu(\vec{n}\times \vec{v})\cdot
\left(\left(1+\frac{3m_1}{4m_2}\right)\vec{S_2}+\left(1+\frac{3m_2}{4m_1}\right)\vec{S_1}\right)\label{Eso2},
\end{equation}
in complete agreement with the result in \cite{spin}. The leading
order spin-spin interaction does not get affected by this new
choice.

\subsection{EOM}

Even though (\ref{Eso2}) is in total agreement with the spin-orbit
potential in baryonic coordinates, the calculation in the
covariant approach does not reproduce the well known fact that the
generalized Lagrangian from which the EOM are derived turns out to
be acceleration dependent \cite{tulc,spin,dam,kidd,will}. Indeed,
(\ref{Eso}) reproduces the gravitational potential in \cite{will}
up to this acceleration dependent piece which does not follow from
a graviton exchange. As we shall show in Appendix E, the solution
to this puzzle lies on the fact that a non canonical algebra
develops which naturally reconciles both approaches. Instead of
following that path here it is instructive to remark there is a
coordinate transformation which leads to a canonical structure.
Not surprisingly this map transforms the covariant choice into the
baryonic one, where it has been explicitly shown there is no need
for an acceleration dependent piece in the action. We can
therefore proceed from the potentials in (\ref{Ess},\ref{Eso2})
and the standard Euler-Lagrange formalism to obtain the EOM within
the baryonic supplementary condition. It can be easily shown that
they are given by (in relative coordinates),

\begin{widetext}
\begin{eqnarray}
\vec{a}\equiv \vec{a}_1-\vec{a}_2 &=&
-\frac{G_NM}{r^2}\vec{n}+\vec{a}_{SO}+\vec{a}_{SS}\label{eom}\\
\vec{a}_{SO}&=&
\frac{G_N}{r^3}\left(3\vec{n}(\vec{n}\times\vec{v})\cdot
\vec{\chi} + 2 \vec{v} \times \vec{\chi}+
3\vec{n}\cdot\vec{v}(\vec{n}\times\vec{\chi})\right)\\
\vec{a}_{SS}&=& -\frac{3G_N}{\mu r^4} \left(\vec{n}(\vec{S}_1\cdot
\vec{S}_2-5\vec{S}_2\cdot\vec{n}\vec{S}_2\cdot\vec{n})+\vec{S}_2(\vec{n}\cdot
\vec{S}_1)+\vec{S}_1(\vec{n}\cdot \vec{S}_2)\right)\\
\frac{d\vec{S}_1}{dt} &=& \frac{G_N}{r^3}\left(\vec{L} \times
\vec{S}_1 \left(2+\frac{3m_2}{2m_1}\right)+\vec{S}_{1}\times
\vec{S}_{2} + 3(\vec{n}\cdot\vec{S}_2)
\vec{n}\times\vec{S}_{1}\right); \;\;\; \frac{d\vec{S}_2}{dt}=1
\leftrightarrow 2,
\end{eqnarray}
\end{widetext}
where we have introduced $M=m_1+m_2$,
$\vec{\chi}=\left(2+\frac{3m_2}{2m_1}\right)\vec{S}_1+\left(2+\frac{3m_1}{2m_2}\right)\vec{S}_2$
and $\vec{L}=\mu r\vec{n}\times \vec{v}$.

From the symmetries of the action we can directly construct the
conserved quantities, in particular the energy. First of all
notice that the spin-orbit {\it force} does not do any work,
namely $\vec{a}_{SO}\cdot \vec{v}=0$, from which we conclude that
the conserved energy is nothing but
\begin{equation}
E=\frac{1}{2} \mu \vec{v}^2 -\frac{G_NM\mu}{r} +
V_{SS}.\label{ene}
\end{equation}

In order to compare with the results in \cite{will} within the
covariant supplementary condition, we restrict ourselves now to
the case of nearly circular orbits to express (\ref{ene}) in terms
of the orbital angular frequency $\omega$ and spin. Taking an
angular average for all quantities we obtain from (\ref{eom}),

\begin{widetext}
\begin{eqnarray}
r(\omega,S) &=&
\frac{M^{1/3}}{\omega^{2/3}}\left(1-\frac{1}{3}{\omega \over
M}\vec{l}\cdot \vec{\chi} - \frac{1}{2} \frac{\omega^{4/3}}{\mu
M^{5/3}}\left(\vec{S}_1\cdot \vec{S}_2 - 3
\vec{l}\cdot\vec{S}_1\vec{l}\cdot\vec{S}_2\right)\right)\\
E(\omega,S) &=&
-\frac{1}{2}(M\omega)^{2/3}\left(1+\frac{4}{3}{\omega \over
M}\vec{l}\cdot \vec{\chi} + \frac{\omega^{4/3}}{\mu
M^{5/3}}\left(\vec{S}_1\cdot \vec{S}_2 - 3
\vec{l}\cdot\vec{S}_1\vec{l}\cdot\vec{S}_2\right)\right)\label{enom},
\end{eqnarray}
\end{widetext}
with $\vec{l}$ the unit vector in the $\vec{L}$-direction. We
therefore conclude that energy as a function of spin and angular
frequency in (\ref{enom}) matches that of \cite{will,kidd}
independently of the spin supplementary condition. As a final
comment let us remark that our results also agree with those of
Buonanno et al. \cite{buo} that appeared after we had completed
our work, where a similar procedure is advocated within baryonic
coordinates. However, the covariant spin supplementary condition
and finite size effects are left undiscussed in \cite{buo}.

\section{Internal degrees of freedom II: Permanent multipole moments}

By now it should be easy to visualize how are we going to include
multipole moments in terms of the $e^I_{\mu}$ fields. Let us
assume for instance the particle has an intrinsic, permanent
quadrupole moment $Q^{IJ}$. In order to couple it to the
gravitational field the following term can be introduced,

\begin{equation}
\int d\tau R_{\mu\alpha\beta\gamma}e^{\mu}_Ie^{\beta}_J
Q^{IJ}u^{\alpha}u^{\gamma}\label{RQ}.
\end{equation}

Notice in fact this is just a generalization of what it has been
done in \cite{nrgr}. In fact, it can be shown (see appendix C)
this quadrupole term is naturally obtained if we include the
non-spherical contribution of the tensor of inertia in the spin
part of the Lagrangian.

From the gauge fixing condition $e^{\mu}_0 \approx p^{\mu}/m$ we
immediately see that the $Q^{0I}$ components do not contribute at
all given that replacing $e^{\mu}_0$ in (\ref{RQ}) give rise to
vanishing terms. We can therefore, as expected, concentrate just
in the spatial components.

\begin{figure}[!t]
\centerline{\scalebox{0.6}{\includegraphics{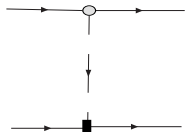}}} \vskip-0.3cm
\caption[1]{Leading order quadrupole-spin one graviton exchange.
The black square represents a quadrupole insertion.}\label{qs}
\end{figure}

It is easy to see that (\ref{RQ}) will naturally reproduce the
quadrupole gravitational energy piece, $\frac{Q_{ij}x^ix^j}{r^5}$,
in the potential. In order to obtain a non trivial contribution we
calculate a correction to the binding energy due to a
quadrupole-spin interaction to leading order. After matching
(\ref{RQ}) into NRGR and using the expression for the Riemann
tensor in the weak gravity approximation, the one potential
graviton exchange in fig.~\ref{qs} gives,

\begin{widetext}
\begin{equation}
V_{QS}=\frac{3G_N}{r^4}\left[({Q_1}^i_i-5 Q_1^{ik}n_in_k)
\vec{n}\cdot(\vec{u}_2\times \vec{S}_2) +
Q_1^{jk}n_k\left((2\vec{u}_2+\vec{u}_1)\times \vec{S}_2\right)_j +
Q_1^{ij}\left({u_1}_j-5(\vec{u}_1\cdot\vec{n})n_j
\right)(\vec{n}\times \vec{S}_2)_i\right],
\end{equation}
\end{widetext}
for the quadrupole-spin potential within the covariant spin
condition, where we have used $Q^{ij}=Q^{ji}$ and the Euclidian
metric ($\delta^{ij}$) to raise and lower indexes. A similar
expression is obtained from the mirror image $1 \leftrightarrow
2$. It is easy to show it corresponds to a 3.5PN contribution for
maximally rotating neutron stars or black holes. Notice that for a
spherically symmetric object ($Q^{ij} \sim \delta^{ij}$)
$E_{QS}\to 0$ as one would have guessed\footnote{This is
reminiscent of the vanishing of the $c_V$ contributions in
\cite{nrgr}.}. Therefore, the coupling is effectively to the
traceless piece of the quadrupole. Higher order multipole moments
are easily handled by similar procedures.

\subsection{Quadrupole radiation}

\begin{figure}[!t]
\centerline{\scalebox{0.6}{\includegraphics{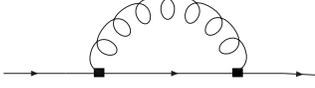}}} \vskip-0.3cm
\caption[1]{Leading order diagram whose imaginary part give rise
to the quadrupole radiation power spectrum. The black boxes are
quadrupole insertions and the curly propagator a radiation
graviton.}\label{im}
\end{figure}

It is instructive to notice that (\ref{RQ}) will directly lead to
the well known quadrupole radiation formula. The leading order
piece will be of the form (for on-shell gravitons),

\begin{equation}
\frac{1}{2m_p}R_{i0j0}Q^{ij}_{TF},
\end{equation}
where $TF$ stands for the traceless piece. By calculating the
imaginary part of fig.~\ref{im} we can immediately obtain (we skip
details which can be found for an identical calculation in
\cite{nrgr}),

\begin{equation}
\mbox{Im}~fig.~\ref{im} =-{1\over  80 m^2_{Pl}}\int {d^3{\bf
k}\over (2\pi)^3 2|{\bf k}|} {\bf k}^4 |Q^{ij}_{TF}(|{\bf k}|)|^2,
\end{equation}
from which the power radiated follows,

\begin{eqnarray}
\nonumber
P &=&  {G_N\over 5\pi T} \int_0^\infty d\omega \omega^6 |Q^{ij}_{TF}(\omega)|^2\\
   &=& {G_N\over 5}\langle \dddot{Q}^{ij}_{TF} \dddot{Q}^{ij}_{TF}\rangle,
\end{eqnarray}
whit dots as time derivatives and the bracket representing time
averaging. This is the celebrated quadrupole radiation formula.

\section{Divergences, non minimal insertions and finite size effects for
spinning bodies}

We are going to discuss here the appearance of divergences and
their consequent renormalization. This will lead us to the study
of higher order terms in the worldline action and their Wilson
coefficients, which will encode the information about the internal
structure of the body. In addition to terms coming from
logarithmic UV divergences, we will encounter power law
divergences whose associated Wilson coefficients do not have scale
dependence. This distinction will turn out to be connected with
tidal deformations vs. self induced effects as well shall see.

\subsection{A cursory first look}

\begin{figure}[!t]
\centerline{\scalebox{0.5}{\includegraphics{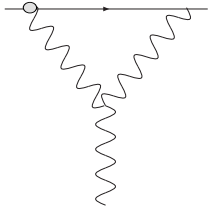}}} \vskip-0.3cm
\caption[1]{Two graviton contribution to the one point function in
effective action with a single spin insertion.}\label{2g}
\end{figure}

Let us study the one point function in the effective action with
spin insertions. Let us start with the diagram shown in
fig.~\ref{2g}. The spin-graviton Feynman rules derived from
(\ref{sg}) differs from mass insertions in two main points: its
tensor structure and its dependence on the graviton momentum.
Fig.~\ref{2g} then contributes for potential gravitons terms
proportional to,

\begin{equation}
\int {d^3{\bf q}\over (2\pi)^3} \left(\frac{\bf q \cdot \bf k,
{\bf q}^2, {\bf k}^2}{{\bf q}^2 ({\bf q} +{\bf k})^2}\right),
\end{equation}
where ${\bf k}$ is the external graviton momentum. None of these
integrals are logarithmically divergent and therefore can be
absorbed as pure counterterms in the original Lagrangian.

\begin{figure}[!t]
\centerline{\scalebox{0.5}{\includegraphics{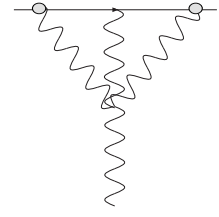}}} \vskip-0.3cm
\caption[1]{Three graviton contribution to the one point function
in the effective action with two spin insertions.}\label{rs}
\end{figure}

Let us concentrate now on the divergent piece coming from diagrams
like in fig.~\ref{rs}. It can be shown that this diagram contain
terms such as (in $d$-dimensions),

\begin{equation}
I({\bf k}) = \int {d^{d-1}{\bf p}\over (2\pi)^{d-1}} {d^{d-1}{\bf
q}\over (2\pi)^{d-1}} {({\bf q}\cdot {\bf k}) ({\bf p}\cdot {\bf
k}) \over {\bf q}^2 {\bf p}^2 ({\bf q} + {\bf p} + {\bf
k})^2}\label{dr},
\end{equation}
as well as integrals with $\bf{q}\cdot\bf{p}$ in the numerator.
These integrals contain power as well as logarithmic UV
divergences. It is clear that these divergences can not be
absorbed into the original Lagrangian since they involve in
principle higher order derivatives of the metric. By general
covariance and parity conservation, there is a limited set of non
zero terms one can build up with the right structure to cancel the
previous divergences. In what follows we will study a possible set
of new insertions in the worldline action which are generated by
renormalization.

\subsection{Non minimal insertions I: Self Induced effects}

We will consider here terms which are not total derivatives and
can not be removed by f.r. Let us proceed systematically. Let us
start with terms linear in Riemann and no further derivatives
acting on external fields. The first non zero terms we can
construct are,

\begin{eqnarray}
{\cal O}^1_{RS^2}&\equiv &
\frac{C^1_{RS^2}}{m_p}R_{\alpha\beta\mu\nu}S^{\alpha\beta}S^{\mu\nu}\nonumber\\
{\cal O}^2_{RS^2}&\equiv &
\frac{C^2_{RS^2}}{m_p}R_{\alpha\beta\mu\nu}S^{\alpha\mu}S^{\beta\nu}\nonumber\\
{\cal O}^3_{RS^2}&\equiv &
\frac{C^3_{RS^2}}{m_p}R_{\alpha\beta\mu\nu}S^{\alpha\gamma}S^\mu_\gamma
u^{\beta}u^{\nu}\label{cs}.
\end{eqnarray}

It is possible to show that they are physically equivalent,
namely, they are proportional up to f.r. removable terms. By
simple inspection it is easy to see they are also similar to the
quadrupole moment insertion in (\ref{RQ}).It is therefore natural
to associate these operators with self induced quadrupole effects
rather than tidal deformations. It is straightforward to show,
after matching into NRGR for potential gravitons, that ${\cal
O}^i_{RS^2}\sim \sqrt{L}v^{2s+2}$ with $C^{i}_{RS^2}\sim 1/m$.
This immediately tell us ${\cal O}^i_{RS^2}$ can not be generated
from a logarithmic UV divergence. It is possible to show
nonetheless that these terms can be generated from power law
divergences (see appendix G). At 2(5)PN these terms generate a
gravitational potential for a maximally rotating (co-rotating)
neutron star or black hole $A$ coupled to a non spinning one $B$
of the form,

\begin{equation}
V_{S^2O} = C^{tot}_{RS^2(A)}
\frac{G_Nm_B}{2r^3}\left(3(\vec{S_A}\cdot\vec{n})^2-\vec{S_A}\cdot
\vec{S_A}\right)\label{s2o},
\end{equation}
with
$C^{tot}_{RS^2(A)}=(-C^1_{RS^2}-\frac{1}{2}C^2_{RS^2}+\frac{1}{4}C^3_{RS^2})_A$.
A spinning particle will tend to deform and therefore generate
multipole (mass) moments which will thus produce a binding energy
term equivalent, as in this case, to a quadrupole
interaction\footnote{It will in principle vary in time and
henceforth radiate. This effect will be naturally taken into
account similarly as we did for the quadrupole.}\cite{poisson}. It
is well known that rotating black holes, or neutron stars, have a
quadrupole moment given by $Q_{bh}=-aS^2/m$ ($G=c=1$), with $m,S$
the mass and spin respectively. For a black hole $a=1$
\cite{thorne}, for neutron stars $a$ ranges between $4$ and $8$
depending on the equation of state of the neutron star matter
\cite{poisson2}. This will give us an straightforward matching for
$C^{tot}_{RS^2(A)}$ in ({\ref{s2o}) which is consistent by
dimensional analysis with what we expect from naturalness
arguments. Furthermore, these coefficients contribute to the one
point function, and thus will show up in the metric solution for a
rotating neutron star or black hole. For the case of a black hole,
the Kerr-Newman spacetime does not have any logarithmic dependence
and therefore, every coefficient associated to a non f.r.
removable term\footnote{As it was noticed in \cite{nrgr}, f.r.
removable terms can in principle show up in the one point
function. However, they can be washed away by a coordinate
transformation. Here we will concentrate in terms which are not
f.r. removable.}, which contributes to the one point function must
be scale independent. Similarly for neutron stars. This provides a
natural characterization for self induced effects. Tidal effects
in the other hand will be then associated to coefficients which do
not contribute to the one point function, and moreover are scale
dependent. Tidally induced effects will be therefore
naturally generated by logarithmic UV divergences.\\

Once self induced spin-multipole moments are included it is no
longer necessary to introduce a non dynamical permanent multipole.
Adding more spin insertions without derivatives will have the same
type of behavior we encountered above, namely the Wilson
coefficients will scale with negative powers of the mass and
therefore they can not be generated from logarithmic UV
divergences. It is indeed possible to show that we can in
principle hook together $n$ spin tensors leading to terms scaling
as $\sqrt{L}v^{4+n(s-1)}$ after matching into NRGR. Given that $n$
could be any number of spin insertions, it appears as if we will
have no predictive power for the case $s=1$. However, it can be
shown that terms with a large number of spin insertions can be
rewritten in terms of interactions with no more than four spin
insertions \cite{regge}. The case of three and four spin
insertions does not modify the leading order expression in
(\ref{s2o}).

\subsection{Non minimal insertions II: Tidal deformations}

Other type of worldline insertions we could in principle generate
are those having derivatives of the Riemann tensor and more spin
insertions. We will need to introduce terms like,

\begin{eqnarray}
& & D_{\epsilon}
R_{\alpha\beta\mu\nu}S^{\epsilon\mu}S^{\nu\beta}u^{\alpha};\;
D^2R_{\mu\nu\alpha\beta}S^{\mu\sigma}S^\alpha_\sigma u^\nu
u^\beta; \label{drs}\\
& & D_{\rho}D_{\sigma}
R_{\mu\nu\alpha\beta}S^{\sigma\mu}S^{\rho\alpha}u^\nu u^{\beta};\;
D_{\sigma}D^2R_{\beta\rho\mu\epsilon}S^{\rho\sigma}S^{\epsilon}_{\gamma}
S^{\beta\gamma}u^{\mu}...\nonumber
\end{eqnarray}

We will concentrate here in tidal effects and therefore in those
terms generated by logarithmic UV divergences. To lowest order it
can be shown we will have $D^2{\cal O}^i_{RS^2}$ (with $i=1,2,3$)
coming from diagrams like fig.~\ref{rs}. We will generically
denote their Wilson coefficients by $C_{D^2}$. Given that these
expressions posses different tensorial structure the RG flow will
naturally decouple. By dimensional analysis it is easy to conclude
that

\begin{equation}
\mu\frac{dC_{D^2}}{d\mu} \sim \frac{m}{m^4_p}.
\end{equation}

As it was pointed out before, these new insertions will not
contribute to the one point function\footnote{For instance
$D^2{\cal O}^i_{RS^2}=0$ on shell.}. However, they will in
principle be observable for more complicated ambient metric,
such as the field produced by a binary companion.\\

Let us power count this effect. After matching into NRGR for
potential gravitons we will get, to leading order,
\begin{eqnarray}
C_{D^2} (\partial^4 H_{\bf k}d^3{\bf k}) S^2 d\tau &\sim &
\frac{m}{m_p^4}\frac{1}{r^4}\frac{v^2}{\sqrt{L}}L^2v^{2s}\frac{r}{v}\nonumber\\
&\sim & \sqrt{L}v^{6+2s},
\end{eqnarray}
which would make it a 4PN contribution for maximally rotating
compact objects. A careful inspection shows however that the
leading piece from these terms goes as derivatives of
$\delta(x_1-x_2)$, which is a contact interaction (the ${\bf k}^2$
piece cancels the propagator). As a consequence, the first long
range interaction coming from $C_{D^2}$, and therefore the lowest
companion induced tidal effect, will scale as $Lv^{8+2s}$, a 5PN
contribution for maximally rotating compact objects (formally at
3PN as shown in appendix B). At this order new terms will also
start to contribute (for instance the third and fourth expressions
in (\ref{drs}). See appendix G for details). The reasoning in
previous sections can be easily extended to the $n$-point function
and higher Riemann insertions.

\section{Conclusions}

In this paper we have extended the formalism initially proposed in
\cite{nrgr} to include internal degrees of freedom like spin as
well as multipole moments. As a first step we have developed in a
suitable fashion the description of spinning bodies in GR to
include a richer tensor structure extending the previous work done
in the realm of special relativity by A. Hanson and T. Regge
\cite{regge}. We have shown that a self consistent action
principle can be implemented and Papapetrou equations \cite{papa}
recovered. Permanent multipole moments are naturally introduced by
adding new degrees of freedom in the worldline action. Using this
formalism we have extended NRGR, its power counting and Feynman
rules with which we have reproduced the well known spin-spin and
spin-orbit effects at leading order \cite{will}. A quadrupole-spin
correction to the binding energy was obtained for the first time
(to my knowledge) as well as the quadrupole radiation formula
recovered. The equivalence between different choices for the spin
supplementary condition was explicitly shown. We have shown
afterwards the appearance of divergences at higher orders in the
PN expansion and its consequent regularization. The type of
divergences are twofold: logarithmic and power law UV divergences.
This distinction was shown to be associated to tidal vs. self
induced effects. Renormalization through the insertion of non
minimal terms in the effective action was implemented and the RG
flow obtained. A finite size cutoff was invoked in the case of
power law divergences and its respective Wilson coefficients set
by naturalness. In the EFT spirit it is likely that all terms
which are consistent with the symmetries will contribute to the
effective action. In fact, self induced spin effects are naturally
expected and the lack of scale dependence just responds to the
fact that it is a 1-body effect on itself due to its proper
rotation which does not get renormalized\footnote{Formally
speaking, self induced effects do not get renormalized as a
consequence of the fact that they are derived from the coupling to
the conserved stress energy and the metric field does not get
renormalized classically.}. A partial matching into the full
theory was accomplished by comparison with known results
\cite{poisson,poisson2}. Self induced effects could in principle
appear at leading orders in the PN expansion in the case of
maximally rotating compact bodies, for which tidal deformations
were shown to first appear at 5PN, although formally at 3PN for
general objects. Within the power of the
EFT, most of the conclusions are based on dimensional grounds without detailed calculations.\\

Several aspects remain still to be worked out. In addition to the
matching calculation and the issue of finite size effects, higher
order corrections are yet to be obtained even though the formalism
is already set and just computational work is needed. Including
dynamical properties for multipole moments as well as back
reaction effects is also to be worked out. Moreover, new
kinematical scenarios, like a 3-body system and the large small
mass ratio case, are currently under study. All these issues,
including the radiative energy loss due to spin will be covered in
forthcoming publications.

\section{Acknowledgments}

I thank Rodolfo Gambini, Walter Goldberger, Ted Newman, Eric
Poisson, Jorge Pullin and Ira Rothstein, for helpful comments and
encouragement. I am particularly indebted to Ira Rothstein, for
discussions during which aspects of this work were clarified, and
a careful reading of the manuscript. I would like to thank also
the theory groups at LSU, PSU and University of the Republic
(Uruguay), for hospitality. This work was supported in part by the
Department of Energy under grants DOE-ER-40682-143 and
DEAC02-6CH03000.

\appendix

\section{Angular-velocity/spin relationship}

By RPI we know the theory has vanishing Hamiltonian and dynamics
is generated by the constraints $C_1,C_2$. It can be also shown
that the Lagrange multiplier associated to $C_1$ has been set to
zero by the condition $\psi \approx 0$ \cite{regge}. Using the
Hamiltonian equations for the tetrad and position we have in the
realm of special relativity,

\begin{eqnarray}
\frac{dx^{\mu}}{d\lambda}&=&[x^{\mu},\xi C_2]_{pb}= 2\xi p^{\mu}
\to
\xi= \frac{(u^2)^{1/2}}{2m}\nonumber\\
\frac{d e^I_{\mu}}{d\lambda}&=&[e^I_{\mu},\xi C_2]_{pb}=2\xi
f'(S^2)S^{\nu\mu}e^I_{\nu} \to\nonumber\\ &\to & \Omega^{\mu\nu} =
\frac{(u^2)^{1/2}}{m} f'(S^2) S^{\mu\nu},\label{osp}
\end{eqnarray}
where $\xi$ is a Lagrange multiplier, $[,]_{pb}$ stands for the
Poisson bracket and $f(S^2)\equiv m^2(S^2)$. By comparison with
(\ref{omegas}) we conclude, by matching to the zero curvature
case,
\begin{equation}
\Omega_{\mu\nu} = (u^2)^{1/2}\frac{f'(S^2)}{m}(S_{\mu\nu} -
\frac{1}{2m}R_{\mu\nu\alpha\beta}S^{\alpha\beta} + ...).
\label{omegasN}
\end{equation}

The Lagrangian dependence of this expression is encoded in the
function $f(S^2)$ defined by (\ref{11},\ref{12}). It is possible
now to construct a Lagrangian ($\bar L$) using all the freedom we
showed previously, that will ensure (\ref{const}) kinematically
\cite{regge}. Such a procedure is therefore preferred given that
the unphysical degrees of freedom are cut off kinematically rather
than cut by hand. As it has been shown in \cite{regge} ${\bar L}$
is however not unique. There is still a remnant freedom of the
form $f'(S^2) \sim A$, with $A$ a constant\footnote{Regge
trajectories are of constant slope.}. We can henceforth set $A$ in
order to recover the well known relationship between
angular-velocity and spin in flat space, namely $S \sim I\Omega$,
with $I$ the moment of inertia. One then solves for $S^{\mu\nu}$
in (\ref{omegasN}) order by order to get (\ref{omegas2}).

One could still argue that the angular-velocity/spin relation
should be obtained directly from $S_M= \left(\frac{\partial {\bar
L}}{\partial \Omega}\right)_M$ rather than using the EOM as we
did. One should however bear in mind that dynamics naturally help
us to power count within an EFT approach. Higher order corrections
are taken into account by insertions in the worldline action
\cite{nrgr}.

\section{Formal power counting}

Here we will comment on the power counting from a formal point of
view without assuming any specific properties of the objects. This
will introduce new parameters which should be adjusted depending
on the constituents. For an object of characteristic length $R$
the spin magnitude for a rotating velocity $v_{rot}$ is $S \sim L
\frac{R}{r}\frac{v_{rot}}{v}$, formally a non perturbative
effects. For maximally rotating, and co-rotating, bodies one gets
$S \sim L \frac{R}{rv}$ and $S \sim L \frac{R^2}{r^2}$
respectively. Nevertheless, it is naturally expected that
$\epsilon\equiv R/r \ll 1$ and a new perturbative parameter is
introduced ($\epsilon \sim v^2$ for neutron stars or black holes).
It is easy to show that the leading spin-orbit effect for
maximally rotating objects scales as $v\epsilon L$, a subleading
contribution formally at 0.5PN. The first long range contribution
from $C_{D^2}$ to the effective action will now scale as $Lv^6
\epsilon^2$ for $v_{rot}=1$. A 3PN effect, effectively at 4PN for
$\epsilon \sim v$. This brings hope to potentially observe these
effects in the future.

\section{Going to all orders}

As we pointed out the {\it unphysical} states can be washed away
kinematically by (\ref{const}) if a suitable Lagrangian ($\bar L$)
is chosen \cite{regge}. Moreover, the leading order spin-graviton
vertex was shown to be Lagrangian independent to leading order.
This however can no be translated to higher orders. Here we will
show how to proceed to obtain the Feynman rules to all orders.\\

By local Lorentz invariance and general covariance we know ${\bar
L}$ is a function of the metric and angular-velocity. We can
rewrite ${\bar L}\equiv {\bar L}(\Omega^{IJ},\eta^{IJ})$ which
shrinks to $\Omega^{IJ}\equiv e^\mu_Ie^\nu_J\Omega_{\mu\nu}$ all
the metric dependence. Within an EFT framework the explicit form
of the Lagrangian given in \cite{regge} is not necessary, since we
can always obtain its NRGR counterpart by expanding ${\bar L}$
around a Minkowski background,

\begin{equation}
{\bar L}= {\bar L}(\Omega^{IJ}_M) + \left(\frac{\partial \bar
L}{\partial \Omega^{IJ}}\right)_M \delta\Omega^{IJ} + ...
\frac{1}{n!}\left(\frac{\partial^n \bar L}{\partial
\Omega^n}\right)_M \delta^n \Omega...,\label{expL}
\end{equation}
where $\delta\Omega^{IJ}(\delta e,h)\equiv
\Omega^{IJ}-\Omega^{IJ}_M$, and $\Omega^{IJ}_M$ defined by
(\ref{omM}). Using that $S_M = I\Omega_M$ on shell, the NRGR
Lagrangian turns out to be (schematically),

\begin{equation}
{\bar L}={\bar L}(\Omega_M)-\frac{1}{2} S_M\delta\Omega -
\frac{I}{2} \delta \Omega\delta\Omega.\label{expL2}
\end{equation}

By expanding $\delta\Omega(\delta e,h)$ in (\ref{expL2}) we will
therefore generate the spin-graviton vertices to all orders in the
weak field limit\footnote{We will show in appendix F that the spin
part of the action can be indeed rewritten in terms of the Ricci
rotation coefficients in a more compelling fashion.}. The next
step to construct the EFT is to match into NRGR using the power
counting rules thus far developed \cite{nrgr}. That is an
straightforward task. The terms in the NRGR Lagrangian for
maximally rotating compact objects are shown in
(\ref{sgnr1},\ref{sgnr15},\ref{sgnr2}) up to 2PN.\\

Notice also that the second piece in (\ref{expL2}) generates
contributions which are not explicitly spin dependent, although
the coupling is proportional to the moment of inertia. For
spherically symmetric objects we will have for instance a term of
the form,

\begin{equation}
\frac{I}{2}~\Gamma^{\mu}_{\alpha\nu}{\Gamma_{\mu\beta}}^{\nu}u^{\alpha}u^{\beta},
\end{equation}
which can be easily shown to be proportional to $R_{\mu\nu} u^\mu
u^\nu$ and henceforth f.r. removable. For non-spherical bodies we
will get,
\begin{equation}
R_{\mu\nu\alpha\beta} e^{\beta}_K e^{\nu}_J I^{KJ} u^\mu u^\alpha,
\end{equation}
with $I^{KJ}$ the inertia tensor defined by $I^{KJ}=\sum_p
m_p\left( {\vec x}^2_p ~\delta^{KJ}- x^J_p x^K_p\right)$ with $p$
labelling the internal structure of the body. Given that the
symmetric piece leads to a Ricci tensor, the only physically
observable contribution will come from the quadrupole piece
$Q^{KJ}=\sum_p m_p x^J_p x^K_p$ as expected. Therefore, by adding
the non-spherical structure into its rotational part we will
automatically account for its internal quadrupole moment
structure. Higher order multipoles do not follow this procedure
and they should be added depending on the physical situation.

\section{Stress energy tensor for spinning objects}

As it was shown by Dixon a spinning particle can be described by
the following stress energy tensor \cite{dixon,mino},
\begin{widetext}
\begin{equation}
T^{\alpha\beta}_{D}= \sum_A \int d\tau
p^{\alpha}u^{\beta}\frac{\delta^4(x^\mu-x^\mu_A(\tau))}{\sqrt{-g}}
-\frac{1}{2}\nabla_\mu\left[\left(S^{\mu\alpha}u^{\beta}+S^{\mu\beta}u^{\alpha}\right)
\frac{\delta^4(x^\mu-x^\mu_A(\tau))}{\sqrt{-g}}\right].\label{bi}
\end{equation}
\end{widetext}

The Papapetrou equations can be recovered as a consequence of
Einstein equations, namely $T_{\alpha\beta;\beta}=0$
\cite{mino}\footnote{ However, they do not decouple using
(\ref{bi}). They can be separately recovered by using the stress
energy tensor proposed by Bailey and Israel \cite{israel}, plus
imposing the symmetry condition $T^{\mu\nu}=T^{\nu\mu}$ by hand
\cite{cho}.}

It can be shown  also that (\ref{bi}) is obtained from our
formalism. By definition the stress energy tensor is defined such
that,

\begin{equation}
\delta S= -\frac{1}{2} \int d^4x \sqrt{-g} T^{\alpha\beta}\delta
g_{\alpha\beta}.
\end{equation}

The variation of the action is in principle tricky due to the
presence of the constraint $e^I_\mu e_{I\nu}=g_{\mu\nu}$. Using,

\begin{eqnarray}
\delta e^I_\mu &=& \frac{1}{2}e^I_\nu g^{\nu\beta}\delta g_{\beta\mu}\\
\delta g^{\mu\nu} &=& - g^{\alpha\nu}g^{\beta\mu}\delta
g_{\alpha\beta},
\end{eqnarray}
we will therefore get,

\begin{equation}
\frac{-1}{2}\int d\tau S^{IJ}\delta \Omega_{IJ} = \frac{-1}{2}
\int d^4x \sqrt{-g} T^{\alpha\beta}_{D(spin)}\delta
g_{\alpha\beta}
\end{equation}
as expected. It is important to remark that, even though at first
sight (\ref{bi}) looks action independent, it utterly depends on
the relationship between spin and angular-velocity which by itself
depends on the particular Lagrangian.

\section{The EOM in the covariant gauge: Non commutative algebra}

As it is known relativistic N-body EOM cannot be derived from an
ordinary Lagrangian beyond the 1PN level, provided Lorentz
invariance is preserved \cite{sanz}. However, the latter does not
follow if the position coordinates are not canonical variables,
and this is exactly what happens once the second class constraints
in (\ref{const}), and the gauge fixing condition $\psi^\mu=0$, are
imposed strongly in the phase space \cite{dirac}. As it is shown
in \cite{regge} the Poisson brackets are modified by the Dirac
algebra \cite{dirac,teit}. The interesting commutation relations
are those of $x^i_a$ and spin, to leading order we have
\cite{regge},

\begin{eqnarray}
{[x^i_a,x^j_a]}_{db}&=&\frac{S_a^{ji}}{m_a^2}\nonumber\\
{[x^k_a,S_a^{ij}]}_{db}&=&\frac{1}{m_a}(S_a^{ki}v^j_a-S_a^{kj}v^i_a)\label{dirac},
\end{eqnarray}
with $a=1,2$, and $db$ stands for Dirac brackets.

It can be easily noticed that our expressions in
(\ref{Eso},\ref{Ess}) for the potentials within the covariant
condition coincides with that of \cite{tulc,dam,will} up to the
acceleration piece. Therefore, all we have to do in order to prove
the equivalence is to explicitly show that the new term derived
from the acceleration part agrees with the extra factor generated
by the non canonical brackets. Let us start with the position
dynamics. It is easy to show from (\ref{dirac}) that the following
new term,

\begin{equation}
\frac{d}{dt}\left(\left[\vec{x}_1,-\frac{G_Nm_2}{r}\right]_{db}\right)=
G_N\frac{m_2}{m_1}\frac{d}{dt} \left(\frac{\vec{n}\times
\vec{S}_1}{r^2}\right),
\end{equation}
will appear into the acceleration of body 1. Adding the piece
coming from the second body it is straightforward to show that the
extra factor is equivalently obtained by adding a term (in
relative coordinates),

\begin{equation}
\frac{\mu}{2M} \vec{v}\cdot\left(\vec{a}\times\left(
\frac{m_1}{m_2}\vec{S}_2+\frac{m_2}{m_1}\vec{S}_1\right) \right)
\end{equation}
into the Lagrangian, where it is understood that wherever the
acceleration appears in higher-order terms one substitutes the
leading order EOM. This agrees with \cite{dam,will} and the
equivalence is thus proven. In addition, it is easy to show there
is a coordinate transformation that leads to a canonical algebra,
to leading order \cite{regge},

\begin{equation}
\vec{x}_a \to \vec{x}_a - \frac{1}{2m_a} \vec{S}_a \times
\vec{v}_a\label{qus}.
\end{equation}

Still missing is the precession of spin. Using (\ref{dirac}) one
can show that the EOM for spin ends up being,
\begin{equation}
\frac{d\vec{S}_1}{dt}= 2(1+\frac{m_2}{m_1})\frac{\mu
G_N}{r^2}(\vec{n}\times\vec{v})\times\vec{S}_1+ \frac{m_2
G_N}{r^2}(\vec{n}\cdot\vec{v}_1)\vec{S}_1.
\end{equation}

It easy to show now that the following PN shift,
\begin{equation}
\vec{S}_a \to (1-\frac{1}{2}\vec{v}_a^2)\vec{S}_a +\frac{1}{2}
\vec{v}_a(\vec{v}_a\cdot\vec{S}_a)\label{shift},
\end{equation}
which jointly with (\ref{qus}) leads to a canonical algebra,
reproduces the well known spin precession (see (\ref{eom})). The
map in (\ref{qus},\ref{shift}) connects the covariant choice with
the baryonic one. The baryonic condition does not preserve Lorentz
invariance, and an acceleration independent Lagrangian
exists\footnote{As shown by Schafer \cite{sch}, substituting the
leading order EOM in the acceleration dependent Lagrangian of
\cite{dam,will} is also equivalent to the map to baryonic
coordinates.}. As a consequence, (\ref{Eso2}) will not be
invariant under the usual linear realization of the Poincare group
\cite{sanz,dam}. Given that (\ref{qus},\ref{shift}) are PN shifts
it is also immediate to conclude that the power counting rules
developed in this paper do not get affected by the new choice.

\section{Spin-graviton coupling revised}

By RPI we know that the Lagrangian must be of the form,

\begin{equation}
L= -p^{\mu}u_{\mu}-\frac{1}{2}S^{\mu\nu}\Omega_{\mu\nu}.
\end{equation}

Here we shall show that the spin part of the action can be
rewritten as,

\begin{equation}
S_{spin} \sim \frac{1}{2}\int d\tau
\omega_{\mu}^{IJ}S_{IJ}u^{\mu}\label{wS},
\end{equation}
with $\omega_\mu^{IJ}$ the Ricci rotation coefficients. Notice
this coupling is generally covariant and RPI by construction. We
could have chosen to add spin this way into the NRGR Lagrangian
given that $S^{IJ}$ can be treated as a constant external source
(see equation (\ref{eq1n})). In fact, the momentum dynamics
(equation (\ref{eq2})) is also recovered following similar steps
as we did before. It is therefore natural to expect that both
formalisms agree which indeed follows almost straightforwardly by
definition. In terms of a tetrad and the Levi-Civita connection,
the Ricci rotation coefficients can be written as \cite{velt},

\begin{equation}
\omega_{\mu}^{IJ}= {{\Gamma^J}_\mu}^I + e_\alpha^J\partial_\mu
e^{\alpha I}.
\end{equation}

Given that (\ref{wS}) is defined on the worldline we can use the
tetrad field transported by the particle. We will thus get for the
spin part of the action (using (\ref{Ome})),

\begin{equation}
u^\mu\omega_{\mu IJ}S^{IJ}= S^{IJ}(-\Gamma_{J I \mu} u^{\mu}+
\frac{de_{\alpha I}}{d\tau}e^\alpha_ J)= S^{IJ}\Omega_{IJ},
\end{equation}
as advertised. Written this way it is clear how spin and gravity
couple to each other, with spin playing the role of a
`gravitational charge' coupled to a connection of a spin 2 field.

\section{Naive power counting}

Here we will schematically show the type of new insertions which
are generated by divergences in the one point function. Let us
start by calculating the effective action with $n_s$ spin and
$n_m$ mass leading order insertions as shown in fig.~\ref{ng}.
This diagram will scale as (after matching into NRGR for potential
gravitons),

\begin{figure}[!t]
\centerline{\scalebox{0.4}{\includegraphics{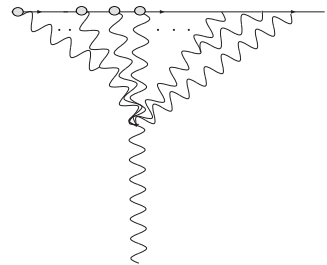}}}
\vskip-0.3cm \caption[1]{A typical contribution to the one point
function in the effective action coming from leading order mass
and spin insertions.}\label{ng}
\end{figure}

\begin{equation}
\frac{1}{m_p^{n_s+n_m-1}}\left(\frac{m}{m_p}\right)^{n_m}L^{n_s}
\frac{v^{sn_s}}{m_p^{n_s}}\frac{v^2m_p}{\sqrt{L}}\frac{r}{r^dv},
\end{equation}
where $r^d$ is introduced for dimensional reasons. By using NRGR
power counting \cite{nrgr} each diagram should scale as $\sqrt{L}$
and therefore,

\begin{equation}
d=2n_s+n_m-1 \to fig.~\ref{ng}\sim \sqrt{L}v^{2d}.
\end{equation}

To consider the type of terms that can be generated by
renormalization in the one point function, and will contribute to
physical observables we need as a necessary condition ${\tilde d}
\equiv d-2 \geq 0$. This however is not sufficient given that
using Bianchi identities it can be shown that contraction of
covariant derivatives with the full Riemann tensor are equivalent
to derivatives of the Ricci tensor and henceforth f.r. removable
terms. We can nonetheless enumerate some cases. Let us concentrate
in logarithmic divergences first. For ${\tilde d}=0$ we have
$n_s=n_m=1$, and it is easy to see there are not any new terms
generated by fig.~\ref{2g}. The case ${\tilde d}=1$ has either
$n_s=2,n_m=0$ or $n_s=1,n_m=2$. None of these diagrams have
logarithmic divergences (for potential gravitons) which could
generate an observable term. In fact, the only observable term
which can be written down with ${\tilde d}=1$, and either
$n_s=1,2$ is the first term in (\ref{drs}), which can be shown to
be a subleading self induced effect.\\ For ${\tilde d}=2$ we have
either $n_s=2,n_m=1$ and fig.~\ref{rs} which leads to the finite
size effects we discussed in the paper, or $n_s=1,n_m=3$ which can
be shown does not generate observable terms. For ${\tilde d}=3$ we
can have $n_s=1,2,3$ plus mass insertions. Some of these diagrams
will contribute observable terms. After matching into NRGR such
terms start out at ${\cal O}(v^{10})$ for maximally rotating
compact objects. The procedure follows systematically with higher
order terms.\\ In addition to logarithmic divergences, it is easy
to see that diagrams like fig. \ref{ng} will also have power law
divergences. For instance, the term ${\cal O}^3_{RS^2}$ can be
generated by diagram fig.~\ref{rs} and its coefficient scale as
$\frac{m\Lambda^2}{m_p^4}$. Assuming a cutoff of order $\Lambda
\sim 1/r_s$, we will have $C^3_{RS^2} \sim
\frac{m}{r_s^2m_p^4}\sim 1/m$ as expected.\\

It is therefore straightforward to conclude from all we have seen
thus far that companion induced finite size effect due to spin
start out at 5PN (formally at 3PN) for maximally rotating compact
objects. For the sake of completeness here are the terms which
will contribute to the potential energy,

\begin{eqnarray}
& & D^2{\cal O}^i_{RS^2},\; D_{\rho}D_{\sigma}
R_{\mu\nu\alpha\beta}S^{\sigma\mu}S^{\rho\alpha}u^\nu u^{\beta},\nonumber\\
& &
D_{\sigma}D_{\epsilon}D_{\gamma}R_{\beta\rho\mu\nu}S^{\beta\sigma}S^{\epsilon\rho}
S^{\mu\gamma}u^{\nu},\;
D^2D_{\sigma}R_{\beta\rho\mu\nu}S^{\beta\sigma}S^{\epsilon\rho}
S^{\mu}_\epsilon u^{\nu},\nonumber\\
& & D^2D_{\sigma}R_{\beta\rho\mu\nu}S^{\rho\sigma}S^{\epsilon\mu}
S^{\nu}_\epsilon u^{\beta},\;
D^2D_{\gamma}R_{\mu\nu\alpha\beta}S^{\gamma\sigma}S^{\sigma\nu}
S^{\mu\beta} u^{\alpha},\nonumber\\
& &
D^2D_{\gamma}R_{\mu\nu\alpha\beta}S^{\gamma\sigma}S^{\sigma\beta}
S^{\mu\nu} u^{\alpha}.
\end{eqnarray}

Other possible terms, like
$D^2D_{\gamma}R_{\sigma\mu\nu\alpha}S^{\gamma\sigma}S^{\nu\alpha}
u^{\mu}$, can be shown to be subleading. The reasoning shown here
can be easily extended to the case of higher order Riemann
insertions with similar conclusions.

\end{document}